# Customer Churn in Mobile Markets: A Comparison of Techniques

Mohammed Hassouna[1], Ali Tarhini[2], Tariq Elyas[3] & Mohammad Saeed AbouTrab[2]

[1] Computing and Information Systems Department, University of Greenwich, United Kingdom

[2] Department of Information Systems, Brunel University London, Middlesex, United Kingdom

[3] Faculty of Arts and Humanities, King Abdulaziz University, Saudi Arabia

Correspondence: Ali Tarhini, Department of Information Systems, Brunel University London, Middlesex, UK. E-mail: ali.tarhini@brunel.ac.uk

**Abstract**

The high increase in the number of companies competing in mature markets makes customer retention an important factor for any company to survive. Thus, many methodologies (e.g., data mining and statistics) have been proposed to analyse and study customer retention. The validity of such methods is not yet proved though. This paper tries to fill this gap by empirically comparing two techniques: Customer churn - decision tree and logistic regression models. The paper proves the superiority of decision tree technique and stresses the needs for more advanced methods to churn modelling.

**Keywords:** customer relationship management, customer churn, data mining, mobile market

## 1. Introduction

Information technology (IT) has become a vital and integral part of every business plan (Tarhini et al., 2014a, b, c; Abbasi et al., 2015; Alenezi et al, 2015; El-Masri et al., 2015; Masa'deh et al., 2015a, b, c; Tarhini et al., 2015a, b, c). Recently, Customer Relationship Management (CRM) has got lot of attentions for nowadays companies where customer retention is considered to be its main factor to be investigated as it focuses on developing and controlling loyal, profitable and lasting relations with customers. Developing successful retention techniques is important for businesses in general, and for mobile operators in particular since they are losing 20% to 40% of their customers each year (Jonathan, Janghyuk, & Lawrence, 2001; Ahn, Han, & Lee, 2006; Seo, Ranganathan, & Babad, 2008; Ordenes et al., 2014; Hu, Shu, & Qiao, 2014, Tarhini et al., 2015d, e; Orozco et al., 2015). Attracting new customers costs a lot in advertising, educating, creating new accounts. Such costs do not exist in the case of retaining existing customers. As a result, keeping an existing customer is five times cheaper than one attracting a new one (McIlroy & Barnett, 2000). Improving customer retention contributes in reducing churn rate from 20% to 10% annually saved about £25 million to the mobile operator Orange (Aydin & Özer, 2005).

Investigate customer churn has been a subject for many methods and techniques including statistical methods and data mining. The superiority of using data mining is well established to investigate customer churn compared with traditional market research surveys (Shaw et al., 2001; Oliveira, 2012; Huang, Kechadi, & Buckley, 2012). Market research surveys based on running questionnaires or interviews voluntarily suffers from a high cost, limited access to the customer population and data self-reporting. Contrary, data mining provides knowledge of whole customer populations based on analysis of their current and historical data. Data mining increasingly becomes a main technique in customer retention to predict future customers' attitude and detect patterns within historical data (Wei & Chiu, 2002; Han, Lu, & Leung, 2012; Liu & Fan, 2014).

Two of the most common data mining techniques are regression analysis and decision tree (Hung, Yen, & Wang, 2006). First, logistic regression predicts the occurrence probability of customer churn by formulating a set of equations, input field values, factors affecting customer churn and the output field (Ahn et al., 2006; Burez & Van den Poel, 2007). Second, decision trees, the most popular type of predictive model (Burez & Van den Poel, 2007; Nie et al., 2011), are used to solve classification problems where the instances are classified into one of two classes (i.e., positives and negatives). However, there is little in the literature that (a) Provides an inclusive view of how decision tree and logistic regression models can be used to analyse and understand customer churn in the mobile telecommunication market; and (b) explores the pros and cons of these techniques.

Addressing these issues, we clarify how to develop decision trees and logistic regression models using real data sets provided by a UK mobile operator. Experimental results and discussion are presented to provide understandings into the complexity of the churn problem. We discuss the pros and cons of techniques in customer retention analysis and suggest directions for future research. This paper is structured as follows. Section 2 revises decision tree and logistic regression modelling techniques. Performance evaluation metrics are described in Section 3. Section 4 introduces modelling experiments results. Section 5 evaluates and compares the performance of the developed models. Lastly, section 6 concludes this paper and points to future work.

## 2. Churn Modelling Techniques

### 2.1 Logistic Regression

Logistic regression is a data mining technique used to predict occurrence probability of customer churn (e.g., Burez & Van den Poel, 2007; Ahn et al., 2006; Miguéis et al., 2012; Faris, Al-Shboul, & Ghatasheh, 2014; Glady, Baesens, & Croux, 2009; Farquad, Ravi, & Raju, 2014). Logistic regression is based on a mathematically-oriented approach to analyse the affecting of variables on the others. Prediction is made by forming a set of equations connecting input values (i.e., affecting customer churn) with the output field (probability of churn). The equations (1), (2) and (3) give the mathematical formulas for a logistic regression model (Miner, Nisbet, & Elder, 2009).

$$p(y = 1 | x_1, \ldots, x_n) = f(y) \tag{1}$$

$$f(y) = \frac{1}{(1+e^{-y})} \tag{2}$$

$$y = \beta_0 + \beta_1 x_1 + \beta_2 x_2 + \cdots + \beta_n x_n \tag{3}$$

Where:

- y is the target variable for each individual j (customer in churn modelling), y is a binary class label (0 or 1);
- $\beta_0$ is a constant;
- $\beta_i$ is the weight given to the specific variable $x_i$ associated with each customer j (j=1,…. m);
- $x_1, \ldots, x_n$ are the predictor variables for each customer j, from which y is to be predicted.

Customer data sets are analysed to form the regression equations. An evaluation process for each customer in the data set is then performed. A customer is can be at risk of churn if the p-value for the customer is greater than a predefined value (e.g. 0.5),

Regression analysis needs to be processed in caution since it may give misleading results when carrying out causality and impact assessment (Cook & Weisberg, 1982; Amin et al., 2014). Multicollinearity resulted from strong correlations between independent variables is a concerning factor in logistic regression models as well (Miner et al., 2009; Hui, Li, & Zongfang, 2013). The existence of strong multicollinearity leads to incorrect conclusions about relationships between independent and dependent variables since it inflates the variances of the parameter estimates and gives wrong magnitude of the regression coefficient. Under certain circumstances, logistic regression can be used to approximate and represent nonlinear systems in spite of it is a linear approach (Cook & Weisberg, 1982; Ren et al., 2014). However, logistic regression needs to be tested and evaluated for the increasing complexity of the mobile telecommunication market (Paul & Cadman, 2002; Su, Shao, & Ye, 2012; Su, 2014).

### 2.2 Decision Tree

Decision trees, the most popular predictive models, is a tree graph presenting the variables' relationships (Burez & Van den Poel, 2007). Used to solve classification and prediction problems, decision tree models are represented and evaluated in a top-down way.

The two phases to develop decision trees are tree building and tree pruning. Starting from the root node representing a feature to be classified, decision tree is built. Selecting a feature can be done by evaluating its information gain ratio. The lower level nodes are then constructed in similar way to the divide and conquer strategy. Improving predictive accuracy and reducing complexity, pruning process is applied on decision trees to produce a smaller tree and guarantee a better generalisation by removing branches containing the largest estimated error rate (Au, Chan, & Yao, 2003). The decision about a given case regarding to which of the two classes it belongs is thus made by moving from the root node to all leaves. Though there are many algorithms for building decision tree, CART, C5.0 and CHAID are those most used (Bin, Peiji, & Juan, 2007).

Decision trees have several advantages. First, they are easy to visualise and understand (Bakır et al., 2009). Second, no prior assumptions about the data are needed since it is a nonparametric approach (Friedl & Brodley, 1997; Rahman et al., 2012). Third, decision trees can process numerical and categorical data. On the other hand, decision trees suffer from some disadvantages. First, its performance is affected by complex interactions among variables and attributes. Second, complex decision trees are very hard to be visualised and interpreted. Third, it suffers from the lack of robustness and over-sensitivity to training data sets (Burez & Van den Poel, 2007; Hassouna & Arzoky, 2011; De, Subbiah, & Vegi, 2014).

## 3. Data Mining Performance Evaluation Metrics

The critical issue in using different churn modelling methods relate to: (a) Efficiently assessing the performance of these methods; and (b), benchmarking and comparing the relative performance among competing models. This section discusses those issues.

### 3.1 Classification Accuracy

The confusion matrix is a tool that can be used to measure the performance of binary classification mode (also called contingency table). A confusion matrix is a visual representation of information about actual and predicted classifications produced by a classification model (Miner et al., 2009, p. 292). Table 1 depicts a confusion matrix for a binary classifier.

Table 1. A confusion matrix for a binary classifier

|  |  | Predicted classes | |
|---|---|---|---|
|  |  | Class=Yes/+/ Churn | Class=No/-/No-churn |
| Actual classes | Class=Yes/+/ Churn | TP (true positive) | FN (false negative) |
|  | Class=No/-/ No-churn | FP (false positive) | TN (true negative) |

Different accuracy metrics *resulted from* the confusion matrix are classification accuracy, sensitivity and specificity. Classification accuracy (CA) is the percentage of the observations that were correctly classified, which can be calculated from the matrix using equation (4) (Vuk & Curk, 2006):

$$CA = \frac{TP+TN}{TP+FP+TN+FN} \quad (4)$$

Classification accuracy is an ambiguous indicator particularly in the case of extreme data. To explain, a base contains 9,990 churn customers and 10 non-churn ones. If a model succeeds to predict that all 10,000 customers are at risk of churn, the accuracy of classification will be 99.9%. The high accuracy rate mistakenly indicates that the model is very accurate in predicting customer churn because the model does not detect any non-churn customers. Predicting churn cases in a correct way is always more important than predicting non-churn cases as the cost of mis-predicting churn is higher than that of mis-predicting non-churn.

### 3.2 Sensitivity and Specificity

Overcoming some of the weakness of accuracy metric, sensitivity and specificity are used. Sensitivity is the proportion of actual positives that are correctly identified. Specificity is the proportion of actual negatives that are correctly identified. The equations are thus:

$$Sensitivity = \frac{TP}{TP+FN} \quad (5)$$

$$Specificity = \frac{TN}{TN+FP} \quad (6)$$

Mobile operators prefer models with high sensitivity rather than models with high specificity because the cost associated with the incorrect classification of churners is higher than the cost associated with the incorrect classification of a non-churner. A compromise between high sensitivity combined with reasonable specificity should be always made so mobile operators can effectively manage their marketing budget to achieve high customer retention.

*3.3 Receiver Operating Characteristic Curve (ROC)*

The Receiver Operating Characteristic (ROC) curve is a depiction of the relations between the true positive rate (i.e., benefits) and false positive rate (i.e., costs), drawn on x and y axis in l leaner scale (Karahoca, Karahoca & Aydin, 2007). ROC represents the relations between the churners ratio correctly predicted as churners, and non-churners ratio wrongly predicted as churners. The ROC provides relative compromises between benefits and costs. The ROC curve consists of points corresponding to prediction results. Figure 1 presents an example of a ROC curve.

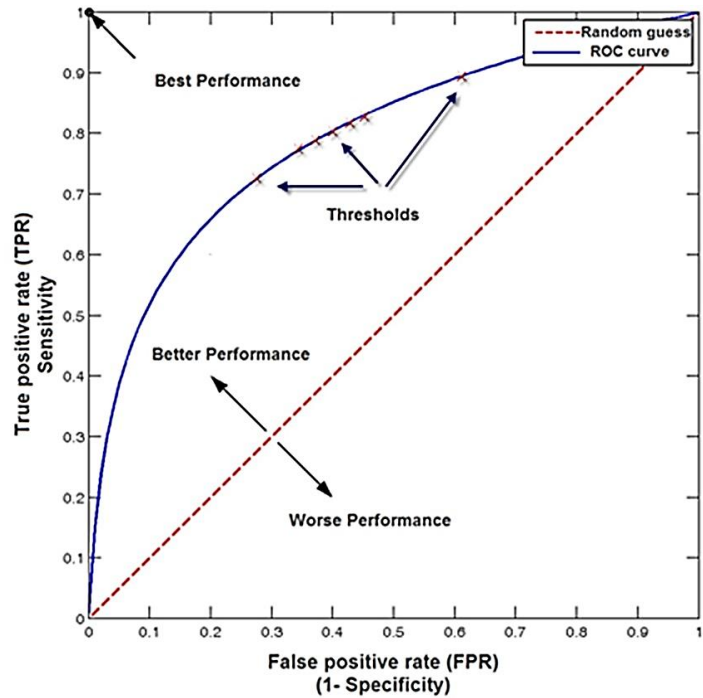

Figure 1. An example of ROC curve

The best performance model is when the ROC curve passes through or close to (0, 1). The model sensitivity and specificity will then be 100% (i.e., no false negatives and no false positive respectively). Some models such as logistic regression produce a score rather than producing binary class decisions (i.e. churn or non-churn). To produce a binary classifier in this case, thresholds are used. If the classifier result is greater than a threshold, the classification class is a churn. Otherwise, it is non-churn. Figure 1 depicts ROC curve for a random predictor represented by the diagonal line that divides the ROC space into two parts represents. ROC curves passing near this line correspond to random guessing classifiers (e.g. classifying by tossing a coin).

Generally, models with ROC curves passing the top left part of the ROC curve perform better. The area under the ROC curve (i.e., called AUC) is used as a performance metric (Karahoca et al., 2007). The AUC value ranges from 0.0 to 1.0. Models perform better when having greater AUC. Moreover, models with AUC value greater than 0.5 perform better than random models because the area under the ROC diagonal line is 0.5.

*3.4 Lift Chart*

Evaluating model performance and determining thresholds, the Lift chart is similar to the ROC curve by yielding high true positive cases. If the total numbers of the classification instances are unknown and the true positive TP rate cannot be computed, the ROC curve is not appropriate and a Lift chart should be used for measuring the accuracy of models (Vuk & Curk, 2006).

In customer churn, the Lift chart gathers customers into deciles based on their churn predicted probability. Each decile shows the predication performance of the model. An instance of lift chart for a customer churn model is shown in Figure 2. The top two deciles arrest about 50% of churners and the top five deciles assert about 90% of churners. Lift charts have become preferred performance evaluation metric used among marketers since it groups customers based on their relative probability to churn. Keeping the marketing cost at a minimum level along with maintaining high customer retention rate can be achieved by targeting customers in the top deciles rather than targeting all of them.

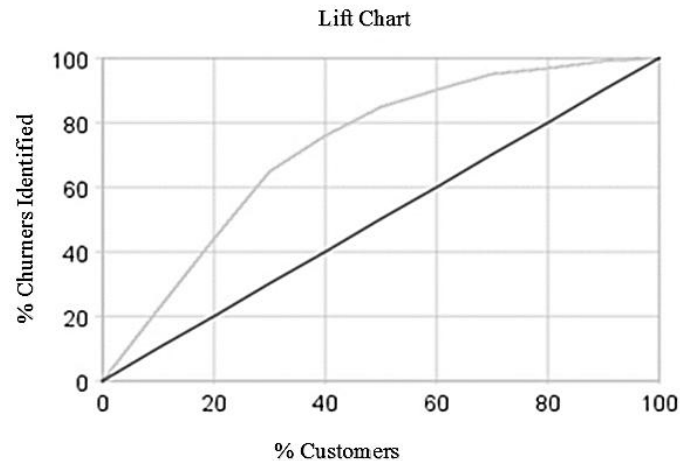

Figure 2. A lift chart of customer churn model

## 4. Modelling Experiments and Results

To test those methods, churn analysis and prediction experiments were used. The ubiquitous data mining methodology CRISP-DM was adopted to investigate customer churn in the telecommunications sector (Chapman et al., 2000; Rudin & Wagstaff, 2014). The CRISP-DM methodology gives comprehensive instructions and procedures for applying data mining algorithms to solve real-world problems. The CRISP-DM model consists of six phases: (1) Business understanding; (2) data understanding; (3) data preparation; (4) modelling; (5) evaluation; and (6) deployment.

Understanding business involves specifying its objectives and correlating them with data mining applications. Understanding data entails data collection, familiarization, exploration, description and quality verification. Data preparation transforming raw data into a suitable format in the sake of applying data mining algorithms includes data cleaning, transformation and reduction. Modelling selects and applies suitable data mining algorithms and identifies their parameters for addressing business problem. The decision about adopting data mining models should be taken based on the evaluation results from a business objective perspective. Deployment is the last phase where the discovered knowledge from the data mining process should be organized and presented in a format that the business can use.

*4.1 Data Sets Description*

This section highlights the processes involved in the data understanding phase. Two sets of data were used to show how traditional analysis techniques may be used to explore churn. The data sets were obtained from a UK mobile telecommunication operator data warehouse. The analysis is based on two datasets of 15,519 and 19,919 customers respectively containing 17 variables. The dependent variable (output variable) simply identifies whether the customer churns or not. The predictor variables (input variables) are in Table 2. These variables are categorised into five groups: Demographics, cost, features/marketing, services usages and customer services. A brief description of the predictor variables is provided by Table 2. Both data sets contained 50% of customers who churned and 50% who stayed with the operator. The data sets also contain customers on 12- and 18-month contracts. The data sets reflect the customer population features under study.

The data sets cover customers' data from March to December, at which the customers took the decision of churn. The data were collected from Month 4 to Month 9 for the new or upgraded 12-month customers. In addition, the data were collected from Month 9 to Month 15 for those on 18 month contracts. The time of the data sets used in this study is illustrated in Figure 3.

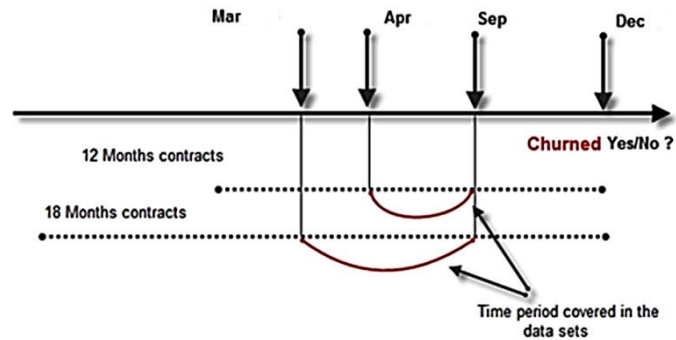

Figure 3. Time window of data sets

Table 2. Description of data sets' variables

| Category | Variable Name | Description |
|---|---|---|
| Demographics | Lifestage_Segment | Subscribers' age stage and gender |
| | Gender | Subscribers' gender |
| | Post_Code | Post code in which subscribers live |
| Cost | Package_Cost | Cost of the package of services chosen by subscribers |
| | Contract_Length | Number of months of the contract |
| | Tenure | Number of months with the present mobile operator |
| Features/Marketing | Tariff | The package of services chosen by subscribers |
| | Device_Desc | Handset model and manufacturer |
| | sales_channel | The first channel where the relationship with the customer was established |
| Usage Level | Q2_bytes | Data usages in the second quarter |
| | Q3_bytes | Data usages in the third quarter |
| | Q2_voice | Voice usages in the second quarter |
| | Q3_voice | Voice usages in the third quarter |
| Customer Services | No_of_Repairs | Number of times handset has been in for repair in a 12-month period |
| | Prob_Handset | Known issues with existing handset |
| | No_of_Complaints | Number of customer complaints regarding billing in a 12 month period |

*4.2 Data Preparation*

The data preparation phase produces the final data sets for models. The main goal of data preparation is to enhance data quality and improve data analysis performance. Data preparation needs to be carried out in a more iterative manner until a conclusive outcome is reached.

In this study, data preparation processes include:

- Imputation of the missing values;
- Discretisation of numerical variables;
- Transformation from one set of discrete values to another;
- Feature selection of the most informative variables;
- New variable derivation.

Imputation process involves replacing missing values with complete information based on an estimate from completed values. Creating new variables from the data is based on discretisation and transformation. Two new variables were formed to measure voice and change in data usage. The performance results of the experiments show that models with selected features outperform those with full feature sets. In spite of the fact that the data sets used in this study have few variables, a feature selection process performed involved three steps:

1) Feature screening by deleting unreliable variables. Two variables (i.e., problem_handset and no_of_complaint) were removed from the feature list because they had very little variation.

2) Feature ranking by sorting the rest of variables based on their importance and correlation with the dependant variable.

3) Feature selection by choosing the most important variables and removing others.

*4.3 Modelling*

In this phase, logistic regression and decision tree techniques were applied and their parameters were calibrated to optimise values. IBM SPSS Modeller (formerly Clementine) was used to build and experiment with those techniques.

*4.4 Logistic Regression Analysis*

A model can be marked 'good' when having a 25% higher classification accuracy rate than the proportional by chance accuracy rate (Costea & Eklund, 2003). The logistic regression analysis was started by computing the proportional by chance accuracy rate based on calculating the proportion of cases for each group (churn and non-churn). The proportions of the non-churn and the churn groups are presented in Table 3. The proportional by chance accuracy rate was then computed by squaring and summarising the proportion of cases in each group. The logistic regression accuracy rates are 25% higher than the proportional by chance accuracy rate. As a result, the classification accuracy criteria are satisfied and the logistic regression model performs better than a random guess.

A relation between the dependent variable (i.e., probability of churn) and combination of independent variables is verified. The independent variables have been added to the analysis and the statistical significance of the model chi-square at step 1 has been checked. The probabilities of the models' chi-squares for both data sets were less than or equal to the level of significance of 0.05. The null hypothesis was accordingly rejected. The existence of a relationship between the independent variables and the dependent variable was thus supported.

Table 3. The proportional by chance accuracy rate

| Data set  | Proportion of churner | Proportion of non-churner | Proportional by chance | Logistic regression model accuracy rate |
|-----------|----------------------|--------------------------|------------------------|-----------------------------------------|
| Dataset 1 | 0.5277               | 0.47220                  | 0.5176                 | 0.681                                   |
| Dataset 2 | 0.499                | 0.5033                   | 0.5023                 | 0.642                                   |

Multicollinearity is a potential issue in logistic regression that is detected by examining the standard errors of the B coefficients. An error larger than 2.0 indicates numerical problems (Miner et al., 2009). None of the independent variables in this analysis had a standard error larger than 2.0. As a result, multicollinearity was not a concern in this study. We present the results of the first data set as the analysis gave very similar results for two data sets.

Figure 4 shows the built model for logistic regression. The stream presents data operations performed on the raw data. Each operation is represented by an icon or node. The nodes are linked together in a stream representing the flow of data through each operation. The classification table for the best performing logistic regression model based on the classification cut-off of 0.5 and 95% confidence intervals (CI for exp (B)) is shown in Table 4. The classification table registers correct and not correct estimations.

Table 4. The logistic regression classification table

| Observed      |      |     | Predicted |      |                    |
|---------------|------|-----|-----------|------|--------------------|
|               |      |     | Flag      |      |                    |
|               |      |     | No        | Yes  | Correct Percentage |
| The Last step | Flag | No  | 5183      | 1294 | 80.0               |
|               |      | Yes | 2175      | 2226 | 50.6               |
|               | Overall Percentage |  |     |      | 68.1               |

Table 5 lists the B parameter, wald statistics, degree of freedom, significance levels and odd ratio. The Wald statistics and the corresponding significance level test the statistical significance of each coefficient (B). If the Wald statistic is significant (i.e. less than 0.05), the variable is significant. Exp (B) values in Table 5 are the results of testing the risk of customer churn. Exp (B) is the predicted change in odds for a unit increase in the corresponding independent variable. If odds ratio is less than 1, a decrease happens in odds. Otherwise, an increase happens in odds. If Odds ratio is 1.0, the independent variable does not affect the dependant variable.

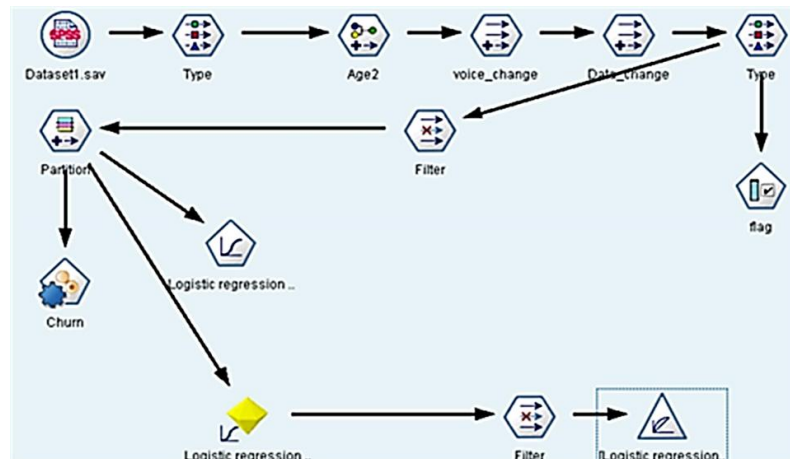

Figure 4. Data stream for the logistic regression model

Table 5. Logistic regression analysis results of the first step

|  |  | B | S.E. | Wald | df | Sig. | Exp(B) |
|---|---|---|---|---|---|---|---|
| Step 1[a] | CNTR_LENGTH | .084 | .009 | 79.717 | 1 | .000 | 1.087 |
|  | Gender | .050 | .050 | .972 | 1 | .324 | 1.051 |
|  | no_of_repairs | -.220 | .072 | 9.282 | 1 | .002 | .803 |
|  | Cost | -.002 | .002 | 1.343 | 1 | .247 | .998 |
|  | billing_queries | -.057 | .062 | .830 | 1 | .362 | .945 |
|  | Tenure | -.026 | .003 | 97.216 | 1 | .000 | .974 |
|  | Coverage | -.289 | .052 | 30.309 | 1 | .000 | .749 |
|  | chang_data_usage | -.159 | .048 | 10.781 | 1 | .001 | .853 |
|  | chang_voice_usage | -.943 | .081 | 133.980 | 1 | .000 | .390 |
|  | bundle_avr_usage | -.106 | .069 | 2.358 | 1 | .125 | .900 |
|  | Constant | .853 | .247 | 11.946 | 1 | .001 | 2.347 |

All of the variables with higher significance level (more than 0.05) are eliminated from the model. Then, the logistic regression model is applied progressively through several iterations to eliminate all non-significant variables. Table 6 shows all the significant variables, which were included in the final logistic model.

Table 6. Logistic regression analysis results of the last step

|  |  | B | S.E. | Wald | df | Sig. | Exp(B) |
|---|---|---|---|---|---|---|---|
| Step 6 | cntr_length | .085 | .009 | 85.360 | 1 | .000 | 1.089 |
|  | no_of_repairs | -.217 | .072 | 9.088 | 1 | .003 | .805 |
|  | tenure | -.027 | .003 | 105.587 | 1 | .000 | .974 |
|  | Coverage | -.292 | .052 | 31.086 | 1 | .000 | .747 |
|  | chang_data_usage | -.162 | .048 | 11.193 | 1 | .001 | .850 |
|  | chang_voice_usage | -.973 | .079 | 152.903 | 1 | .000 | .378 |
|  | Constant | .735 | .217 | 11.461 | 1 | .001 | 2.085 |

*4.5 Decision Tree Analysis*

Customer churn behaviour was investigated by creating a decision tree model. Three decision tree algorithms (CART, C5.0 and CHAID) were utilised to evaluate their performance. The accuracy rates of the decision tree were 68.57% with respect to CART, 68.83% with respect to the CHAD and 70.25% with respect to the C5. The results of C5 are discussed in more detail as it has the best predictive performance. The decision tree constructed using C5 model is shown in Figure 7. The most important independent variables are cntr_length, chang_voice_usage and tenure. Ranking independent variables based on their importance has been shown to be consistent with the logistic regression results.

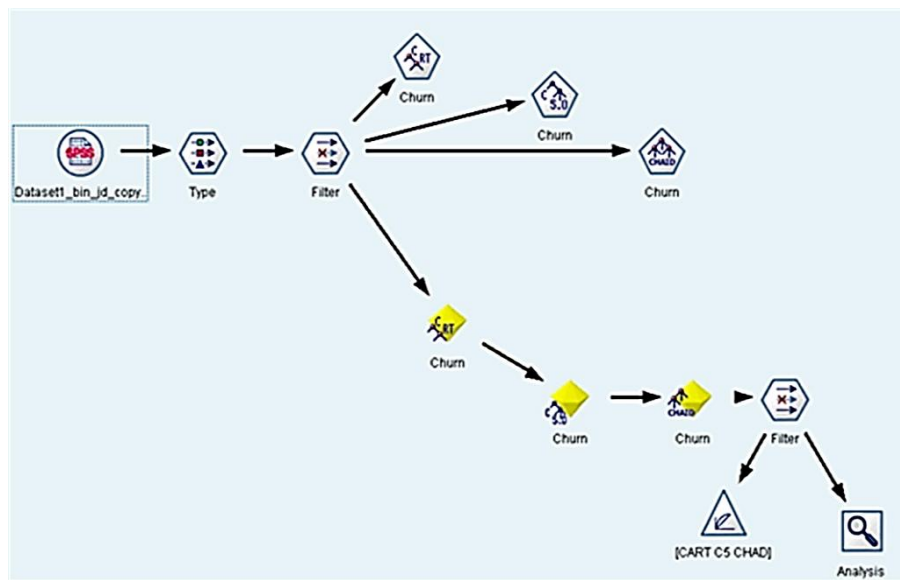

Figure 5. Data stream for the decision tree model

**5. Model Evaluation and Discussion**

From a data-analysis perspective, the models presented in the last section were chosen based on the quality of their outputs. In the evaluation phase, a performance comparison of the different modelling techniques is carried out. A decision on which modelling technique to be adopted for deploying the final customer churn model should be taken at the end of this phase. AUC, ROC curve, top-decile and overall accuracy are used here to compare the performance of the modelling techniques. Based on the evaluation processes, the limitations of data mining in customer churn analysis are discussed.

Figure 6 depicts the ROC curve for three decision trees and one logistic regression models. Table 7 presents the three evaluation metrics adopted in this study. Figure 6 and Table 7 proves that decision trees and C5 outperform the logistic regression models. These results are consistent with some prior literature (e.g., Bakır et al., 2009), having in mind that achieving fair comparison is difficult because of the difference in the data sets used. This study suggests decision tree analysis as a potentially valuable tool for churn prediction based on the evaluation results and the data sets used.

Table 7 shows that C5 model outperforms all other models, including the Logistic Reg. (2) model developed by a data analytics team working for the mobile operator. C5 model in Figure 7 shows a lift value of 1.598 at 30 percentile. To explain, C5 model can capture about 16% of the churners in the top 30% of a sorted list of churned customers.

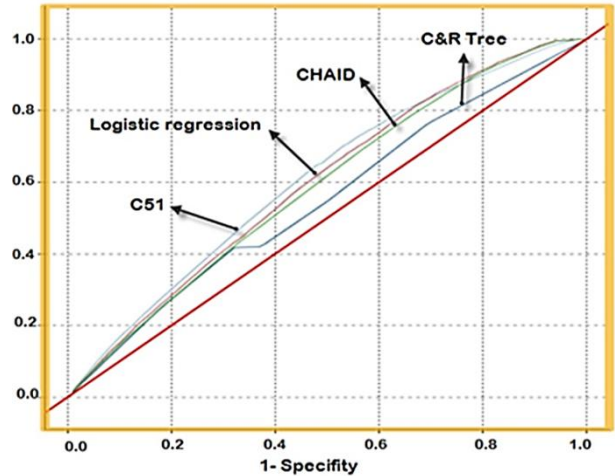

Figure 6. The ROC curve of decision trees and logistic regression models

Table 7. Model evaluation metrics

| Model | Lift (top 30%) | Area under curve | Overall accuracy |
| --- | --- | --- | --- |
| C5 | 1.598 | 0.763 | 70.25 |
| CHAD | 1.584 | 0.710 | 68.83 |
| CART | 1.409 | 0.603 | 68.57 |
| Logistic Reg. | 1.218 | 0.723 | 68.1 |
| Logistic Reg.(2) | 1.4 | - | 67.5 |

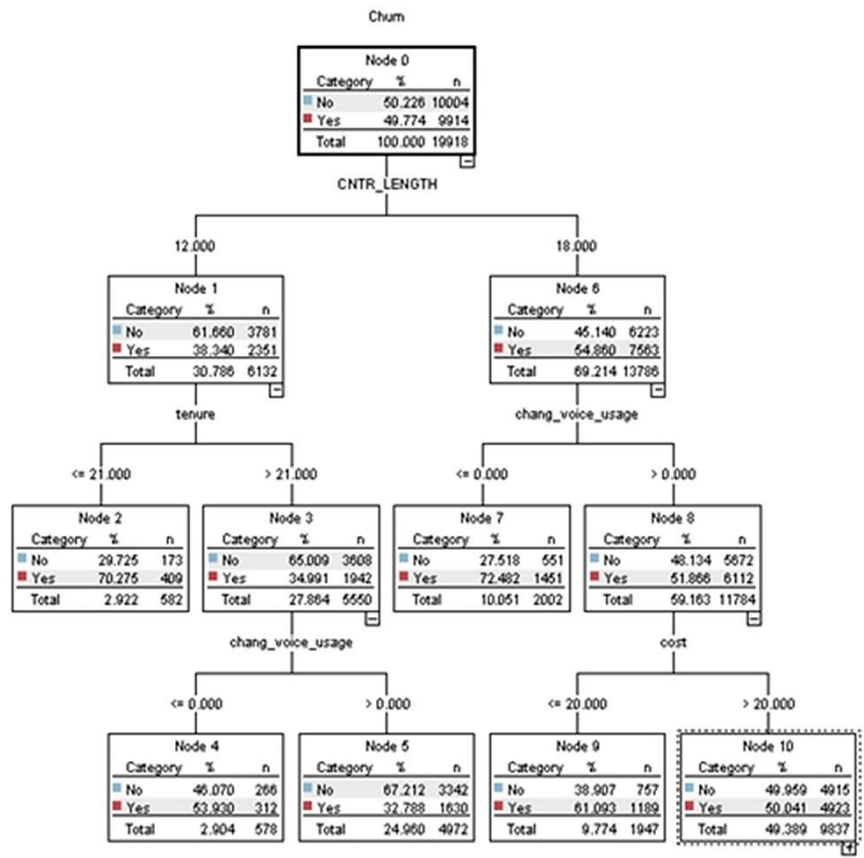

Figure 7. The C5 decision tree model

## 6. Limitations of Data Mining in Customer Retention Analysis

The results the data mining models are reported in this paper prove that prediction accuracy can be improved further by fine tuning model parameters. Specifically, the experimental results show that the prediction accuracy of the models developed has improved accuracy by 1-3% compared to that achieved by the mobile operator. Although data mining is an advanced customer churns analysis tools, the following potential limitations to its use exist:

1) There is no direct value explanation of these patterns. Skilled data mining professionals are needed to provide such an explanation (Friedl & Brodley, 1997).

2) Although data mining links between variables and customer behaviours, it fails to identify the causal relationships between variables (Cook & Weisberg, 1982). Decision trees can be used in some cases to infer causal relationships.

3) Data mining models have a relatively short expiry life. The mobile market faces new technologies on a daily bases. As a result, historical data become less valuable for predictions.

4) Churn relates to complex interactions within population. Examining all the factors affecting customer churn simultaneously and jointly by building a model is not applicable.

5) The level of the analysis in the data mining models decreases the ability to capture the heterogeneity of the customers. New techniques are thus required to support customer churn analysis to bypass the cons of the data mining techniques.

Table 8 lists the limitations of using data mining tools in customer churn analysis and describes the general requirements to overcome them. Agent-Based Modelling and Simulation (ABMS) technique is among the techniques that have the potential to advance the customer churn analysis and overcome some of the limitations of data mining (Paul & Cadman, 2002). ABMS is an emerging means of simulating behaviour and examining behavioural consequences. ABMS is a computational model for simulating the actions and interactions of autonomous individuals in a network, with a view to assessing their effects on the system as a whole. In outline, agents represent customers and agent relationships represent processes of agent interaction.

Table 8. Data mining limitations in customer churn analysis

| No. | Description of the limitation | How to overcome it |
| --- | --- | --- |
| 1 | Explanation of the value and the significance of the patterns and relationships necessitate skilled data mining professionals | Provide an easy and intuitive way to explain the value and the significance of relationships between variables |
| 2 | Low efficiency of identifying causal relationship | Enhance the ability of identifying causal relationships |
| 3 | Short expiry date and the dependency on data quality | Offer tools for long-term planning, and minimise the dependency on data quality |
| 4 | Neglecting customers' interactions | Include customer interaction in customer churn analysis |
| 5 | Cannot captures the heterogeneity of customers | Offer tools that take account of the heterogeneity of customers |

## 7. Conclusion

Many mobile operators have used data mining techniques such as regression and decision trees to overcome churn problems. The experimental results conducted in this study show that the decision tree model outperforms all the logistic regression models examined, including the model developed by a data analytics team working in the mobile operator. The decision tree model produces a lift value of 1.598 at 30 percentile compared with 1.4 at the same percentile produced by the data analytics team. For this problem and for similar data sets, therefore, the conclusion is that decision trees are a preferable technique for investigating customer churn. More reflectively, although data mining provides very useful insight into customer churn, limitations are apparent in relation to significance, causality, data evolution, model complexity and aggregation. In order to overcome these limitations new ways of understanding and exploring data may be helpful - Agent Based Modelling Simulation (ABMS) provides one such avenue for more complex investigations on customer retention.